\documentclass[twocolumn,prl,superscriptaddress,showpacs,byrevtex]{revtex4}
\usepackage{epsf,psfig,epsfig,graphicx,amssymb}

\begin{document}
\title{Observation of Sommerfeld precursors on a fluid surface}
\author{\'Eric Falcon}
 \email[Corresponding author. Email address: ]{Eric.Falcon@ens-lyon.fr}
 
\homepage{www.ens-lyon.fr/~efalcon/}
\author{Claude Laroche}
\affiliation{Laboratoire de Physique, \'Ecole Normale Sup\'erieure de Lyon, UMR 5672, 46, all\'ee d'Italie, 69 007 Lyon, France}
\author{St\'ephan Fauve}
\affiliation{Laboratoire de Physique Statistique, \'Ecole Normale Sup\'erieure, UMR 8550, 24, rue Lhomond, 75 005 Paris, France}

\date{\today}

\begin{abstract}  
We report the observation of two types of Sommerfeld precursors (or forerunners) on the surface of a layer of mercury. When the fluid depth increases, we observe a transition between these two precursor surface waves in good agreement with the predictions of asymptotic analysis. At depths thin enough compared to the capillary length, high frequency precursors propagate ahead of the ``main signal'' and their period and amplitude, measured at a fixed point, increase in time. For larger depths, low frequency ``precursors'' follow the main signal with decreasing period and amplitude. These behaviors are understood in the framework of the analysis first introduced for linear transient electromagnetic waves in a dielectric medium by Sommerfeld and Brillouin \cite{Brillouin60}.

\end{abstract}

\maketitle

One feature of linear wave propagation in a dispersive medium is the existence of precursors (or forerunners). This terminology traces back to the fact that they generally arrive sooner than the ``main'' signal. This transient response is due to the propagation of the fastest high frequency components of the initial spectrum. Although predicted as early as 1914 by Sommerfeld and Brillouin~\cite{Brillouin60}, experimental observations of forerunners are very few and only qualitative, mainly dealing with electromagnetic (e.m.) waves in dielectric medium in the microwave \cite{Pleshko69} or optical \cite{Aaviksoo91} frequency range. Such Sommerfeld forerunners have also been predicted in various dispersive media such biological \cite{Albanese89} or viscoelastic \cite{Hanyga02} ones, and have been recently shown to be linked to the non violation of Einstein causality during superluminal light pulse propagation in the region of anomalous dispersion \cite{Mojahedi00}. 
However, for waves in fluids, the observation of forerunners is still lacking despite some effort performed with acoustic waves in superfluid $^3$He \cite{Varoquaux86} or with pressure waves in fluid-filled collapsible tubes \cite{Kececioglu81}. We report here the first observation of two types of Sommerfeld forerunners, which can coexist or not, on a thin layer of mercury. The non-monotonous dispersion relation of waves on the surface of a fluid lead to a rich variety of such transient wave phenomena. In the shortwavelength limit, when capillary effects are dominant (the analogue to ``anomalous'' dispersion \cite{anormale} for e.m. waves), only the fast high frequency components of the initial excitation are observed arriving before the main pulse, the so-called Sommerfeld precursor for e.m. waves. This transient is characterized by small amplitude  and rapid oscillations (with respect to the main signal). For longer wavelengths, when gravity is no longer negligible (``normal'' dispersion), a slower low frequency precursor is also observed (no analogue exists for e.m. waves).   The forerunners are found in good agreement with the predictions of asymptotic analysis based on the ``stationary phase method''. We note that this study also allows us to connect the forerunner concept of electromagnetic waves to the well-known hydrodynamic transient surface wave phenomena \cite{Kranzer59}, and their applications to submarine eruptions \cite{Smith95}. 

\begin{figure}[ht!]
\epsfig{file=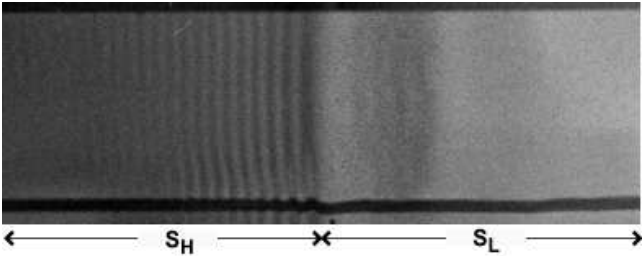, width=8.5 cm}
 \caption{\label{fig:photo}Photograph of typical wave precursors at the surface of mercury (seen from above). Pulse fronts are located on the left.  Fluid depth is $h =3.7$ mm (Bo=0.22), and the full vertical scale corresponds to the 7 cm canal width.}
\end{figure}

\begin{figure}[ht!]
\begin{tabular}{c}
\epsfig{file=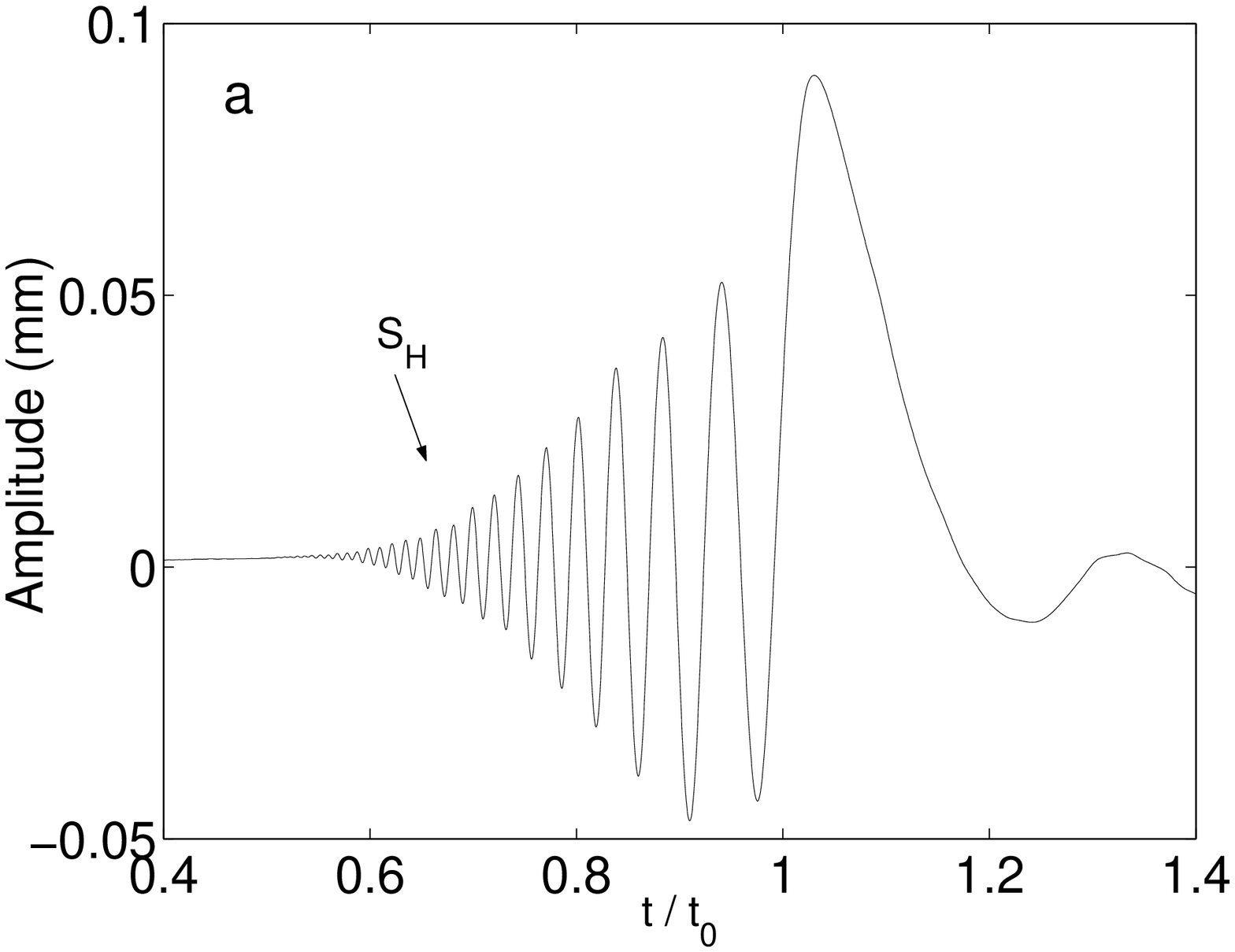, width=8.25 cm}
\\
 \epsfig{file=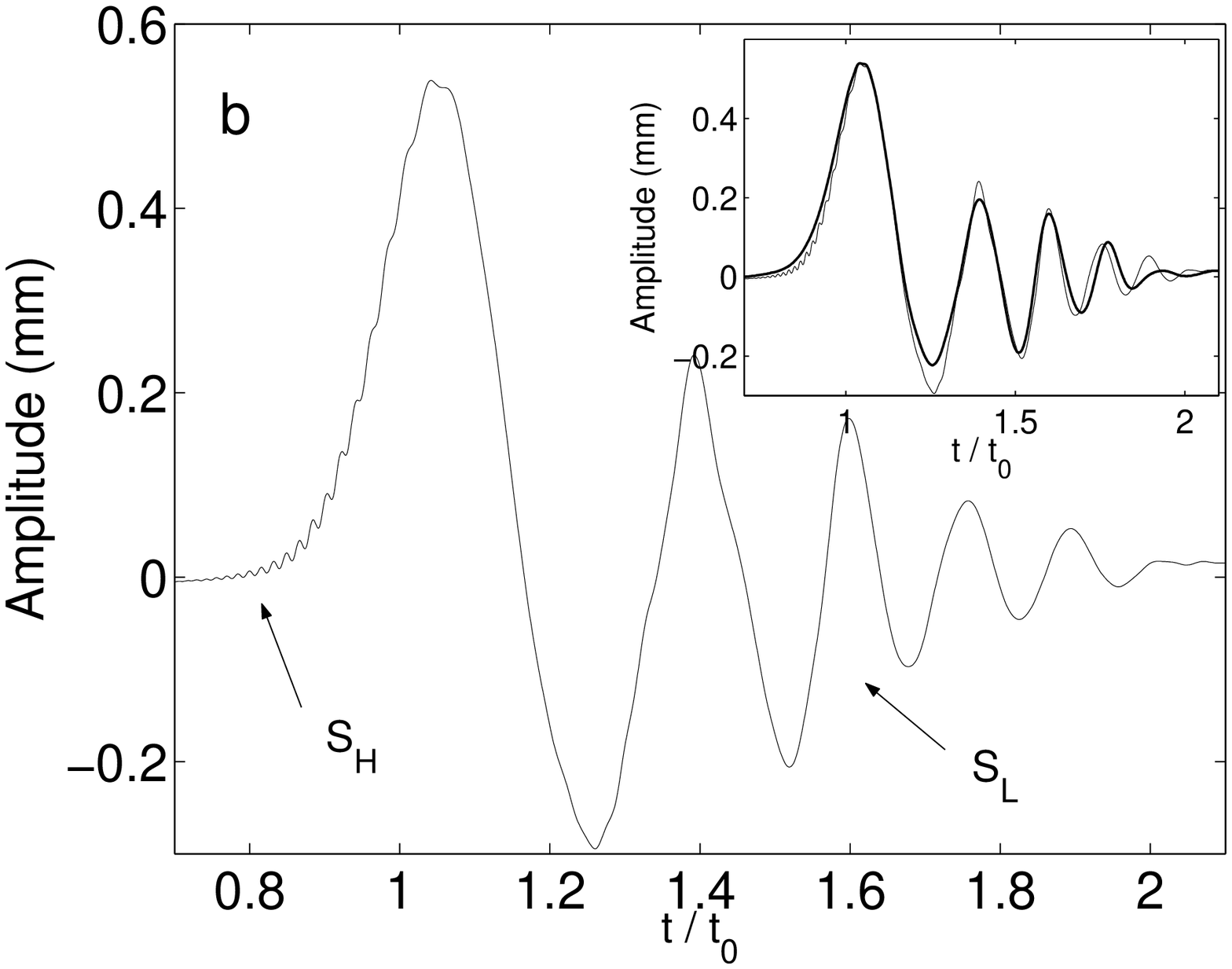, width=8.25 cm}
\\
\epsfig{file=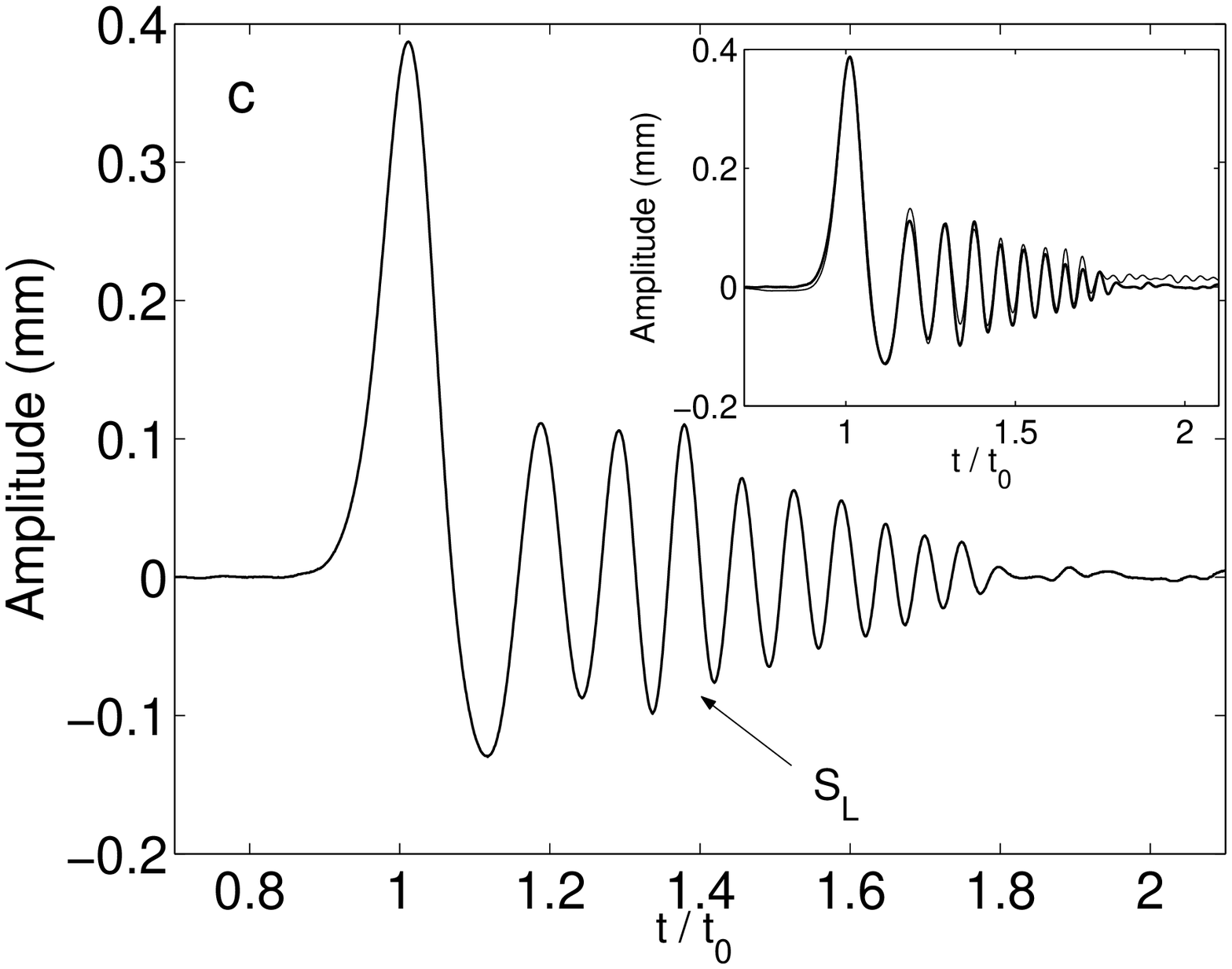, width=8.25 cm}
\\
\end{tabular}
 \caption{\label{fig:formeBF}  Free-surface profiles of Sommerfeld precursors as a function of dimensionless time ($t_0\equiv x/\sqrt{gh}$): (a) Type $S_{H}$ for $h=2.12$ mm at $x=0.2$ m, (b) Type $S_{H}$ and $S_{L}$ for $h=7.2$ mm at $x=0.2$ m, and (c) Type $S_{L}$ for $h=7.2$ mm, $x=0.6$ m.  Inductive (thick line) or optical (thin line) measurements. Pulse fronts are located on the left. Insets show a comparison between both techniques.}
\end{figure}

The experimental setup consists of a 1.5 m long horizontal PMMA channel, 7 cm wide, filled with mercury up to a height, $h$, where: $2 \lesssim h \lesssim 14$ mm. $h$ is measured to a precision of $\pm$ 0.02 mm by means of a depth gauge using a micrometric linear positioner. The properties of the fluid are, density, $\rho = 13.5$ 10$^{3}$ kg/m$^3$, dynamic viscosity, $\eta=1.5$ 10$^{-3}$ Ns/m$^2$ \cite{Handbook}, and surface tension $\gamma=0.4$ N/m \cite{Falcon02}. 
Surface waves are generated by an impulsional excitation provided by the horizontal motion of a rectangular plunging PTFE wavemaker driven by an electromagnetic vibration exciter.  They are generated 10 mm inward from one end of the channel and the local displacement of the fluid in response to this excitation is measured simultaneously by a nonintrusive inductive sensor and by an optical technique \cite{Falcon02}. 
The inductive sensors, 3 mm in diameter, are suspended perpendicularly to the fluid surface at rest. The linear sensing range of the sensors allows distance measurements from the sensor head to the fluid surface up to 2.5 mm with a 5 V/mm sensitivity. 
An optical determination of the local slope of the surface is also performed. Using a position sensitive detector, we have recorded the deflection of a laser beam by the surface wave; the computation of the surface elevation from the optical signal is in excellent agreement with the direct inductive measurement of the shape of the wave [see inset of Fig.\ \ref{fig:formeBF}(b-c)]. Both sensors are mounted on a horizontal linear positioner at a distance $x$ from the wavemaker, $0 < x < 1.2$ m. The optical and inductive methods are complementary as the spatial resolution and the sensitivity of the optical technique are higher, but the inductive technique provides a direct measurement of the surface displacement and does not require signal processing. Both techniques are not limited by their response time in the frequency range of surface waves.
The choice of mercury has been motivated by the possible use of the inductive measurement technique and also because of its low kinematic viscosity which is an order of magnitude smaller than that of water, thus strongly reducing wave dissipation.
Photographs of the free-surface of mercury are made by means of a camera mounted above the center of the channel. A typical pattern is displayed in Fig.\ \ref{fig:photo} showing the high frequency Sommerfeld precursors, $S_{H}$, ahead of the main signal followed by low frequency ones $S_{L}$. 

\begin{figure}[ht!]
\epsfig{file=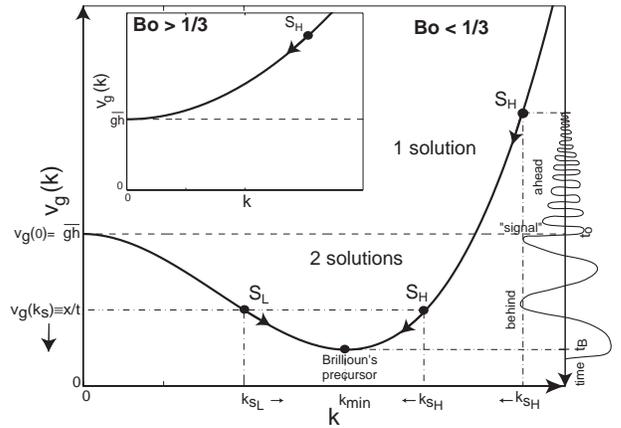, width=8 cm}
 \caption{\label{fig:vitgroup} Group velocity vs wave number for gravito-capillary waves. At fixed $x$ and $t$, the stationnary phase method [$v_g(k_s) \equiv x/t$] leads to solutions with wave number $k_s$. For $h> h_c$ (Bo $<1/3$):  at small $t$, high frequency Sommerfeld precursors, $S_{H}$, are predicted ahead of the ``main signal'' with a period which increases with time ($k_{s_{H}}$ decreases). When $t$ is large enough ($x/t<\sqrt{gh}$) $S_{H}$ switches to a low frequency Sommerfeld precursor, $S_{L}$, with a decreasing period ($k_{s_{L}}$ increases). A Brillouin precursor corresponds to a constant period ($k_{min}$). The typical signal amplitude vs time is sketched on the right. The case $h< h_c$ (Bo $>1/3$), where only the fast $S_{H}$ solution exists, is displayed in the inset.}
\end{figure}

To understand this behavior, one can introduce the dispersion relation for surface waves, neglecting dissipation, 
\begin{equation}
\omega = \sqrt{\left(gk+\frac{\gamma}{\rho}k^3\right)\tanh{kh}}{\rm \ ,}
\label{disp}
\end{equation}
with $\omega$ the wave pulsation, $k$ the wave number and $g$ the acceleration of gravity. From Eq.(\ref{disp}), we can define the capillary length, $l_c \equiv \sqrt{\gamma/\left(\rho g\right)}$, and the Bond number, ${\rm Bo} \equiv \left(l_c/h\right)^2$. In the long wave length approximation or ``shallow water" limit ($kh << 1$), Eq.\ (\ref{disp}) may be expanded, and the group velocity becomes,
\begin{equation}
v_g \equiv \frac{d \omega}{dk}=\sqrt{gh}\left(1-a_2k^2h^2+\frac{a_4}{4}k^4h^4\right){\rm \ ,}
\label{dispbis}
\end{equation}
with $a_2=(\frac{1}{3}-{\rm Bo})$, and $a_4=(19/90 - {\rm Bo}/2-{\rm Bo^2}/3)$.
Equation (\ref{dispbis}) shows that a minimum of the group velocity exists only when $0\le {\rm Bo} < 1/3$,  ${\rm Bo} = 1/3$ corresponding to a critical depth, $h_c$. Figure \ref{fig:vitgroup} shows a qualitative sketch of the wave group velocity as a function of $k$, for ${\rm Bo}$ above and below $1/3$. As explained hereafter, the existence of this minimum has a strong influence on the dispersion of an initial disturbance. 

Generally speaking, if we look at the propagation along $Ox$ of an initial perturbation $\zeta_0(x)$ in a dispersive medium, $\zeta(x,t)$ (e.g. the free-surface deformation) is formally given by the Fourier Integral $\zeta(x,t) = \int_{-\infty}^{+\infty}\hat{\zeta}_o(k)e^{i\phi t}dk$, where $\phi \equiv kx/t- \omega(k)$, with $\omega(k)$ solution of Eq.\ (\ref{disp}) and where $\hat{\zeta}_o(k)$ is the Fourier transform of $\zeta_o(x)$ \cite{Whitham74}. The method of the stationary phase \cite{Jeffreys56,Whitham74} is particulary useful for the asymptotic behavior of these integrals: for both large $x$ and $t$ ($x\gg L\simeq 2$ cm and $t\gg T\simeq 100$ ms, the size and the duration of the initial disturbance), with $x/t$ held fixed (for an observer traveling at this given speed), the main contribution to the integral is from the neighborhood of stationary points $k_s$ such that
\begin{equation}
\left. \frac{d\phi}{dk}\right|_{k_s}=0{\rm \ , i.e. \ } \left. \frac{d\omega}{dk}\right|_{k_s}  \equiv v_g(k_s)=x/t {\rm \ ,}
\label{statio}
\end{equation}
the other components oscillate too rapidly in order to contribute. One can apply graphically this stationary phase method to the dispersion relation of Eq.\ (\ref{disp}), as shown in Fig.\ \ref{fig:vitgroup}. At a fixed point of observation $x$, the main contribution to the surface deformation $\zeta(x,t)$ at any time $t$ results from the points on the group velocity curve equal to $x/t$ (see Fig.\ \ref{fig:vitgroup}). For $h> h_c$ (Bo $<1/3$), three types of precursor are predicted: the fastest signal is the high frequency Sommerfeld precursor, $S_{H}$, (from the capillary branch) ahead of the ``main signal'' (arriving at $t_0$ with velocity $x/t_0\equiv \sqrt{gh}$); 
then, the low frequency Sommerfeld precursor, $S_{L}$ (from the gravity branch), and finally, at $t_B$, the so-called Brillouin precursor (minimum of the $v_g(k)$ curve, i.e. $\phi''=0$) as defined in the framework of e.m.  waves. For $h< h_c$ (Bo $>1/3$), only the fast $S_{H}$ solution exists (see Fig.\ \ref{fig:vitgroup}). Fig.\ \ref{fig:vitgroup} also shows that $S_{H}$ has an increasing period as time goes on (contrary to $S_{L}$), whereas the Brillouin precursor has a constant period. This rich variety of such transient waves comes from the non-monotonous dispersion relation of Eq.\ (\ref{disp}). Finally, returning to the general case, one can have access to the disturbance profile $\zeta(x,t)$ which is given by the above Fourier Integral. A Taylor serie expansion of $\phi(k)$ in the neighborhood of $k_s$ leads to \cite{Jeffreys56,Whitham74}
\begin{equation}
\zeta(x,t) \simeq \sum_{k_s} \hat{\zeta}_0(k_s)\sqrt{\frac{2\pi}{t \left| \left. dv_g/dk\right|_{k_s}\right| }} \cos{[\phi(k_s)t \pm \pi/4]} {\rm \ .}
\label{solution}
\end{equation}
Thus, the shape of the precursor notably depends on the Fourier transform of the initial disturbance, $\hat{\zeta}_o(k_s)$, whereas the temporal evolution of the precursor period depends only of the dispersion relation through $\phi(k_s)$. Therefore, the measured period can be easily compared with the theoretical prediction, whereas the prediction of the experimental profile requires the precise knowledge of the initial conditions. Note that, at the minimum of the group velocity where the Brillouin precursor is predicted, Eq.\ (\ref{solution}) is no longer valid, and the correct asymptotic behavior is found by keeping higher orders in the Taylor series for $\phi(k)$ \cite{Jeffreys56,Whitham74}.

We have performed an experimental study of precursor waves for a fluid layer height in the range $2.12 \leq h \leq 13.75$ mm; thus  $0.02 \leq {\rm Bo} \leq 0.67$. For mercury ($l_c= 1.74$ mm), the critical case ${\rm Bo}=1/3$ corresponds to $h_c \approx 3$ mm. A horizontal impulsion is imposed to initiate the surface wave. In response to this initial disturbance, the free-surface profile is recorded at a fixed distance from the wave generator, and displayed in Fig.\ \ref{fig:formeBF}(a) for $h< h_c$ (Bo $>1/3$) in Fig.\ \ref{fig:formeBF}(b) for $h> h_c$ (Bo $<1/3$). Figure\ \ref{fig:formeBF}(a) shows the high frequency Sommerfeld precursor which propagates at least two times faster than the main signal ($t/t_0 \gtrsim 0.5$), with an increasing period (defined as the time between two successive extrema) as graphically predicted in the inset of Fig.\ \ref{fig:vitgroup}. For a deep enough fluid (Bo $<1/3$) this fast precursor coexists, as shown in Fig.\ \ref{fig:formeBF}(b), 
with low frequency Sommerfeld precursor which appears behind the main signal, with a decreasing period as graphically predicted in Fig.\ \ref{fig:vitgroup}. Moreover, the amplitude of the fast precursor, $S_H$, is much smaller than $S_L$ (see vertical scales in Fig.\ \ref{fig:formeBF}(a-b)). This can be understood by assuming an initial normalized hump of section $s$, $\zeta_0(x)=s^2/(s^2+x^2)$, whose Fourier transform is $\hat{\zeta}_0(k)\sim e^{-ks}$. $e^{-ks}$ being maximum for $k=0$, the precursor $S_L$ with the larger spatial scale has the larger amplitude. For the same depth of fluid as in Fig.\ \ref{fig:formeBF}(b), Fig.\ \ref{fig:formeBF}(c) shows the profile of waves after 0.6 m of propagation. The fastest precursor has disappeared, and only the contribution due to the gravity branch, i.e. $S_L$, is observed.  This is linked to the arguments given above, and also to viscous dissipation.
The inset of Fig.\ \ref{fig:formeBF}(b-c) shows good agreement between optical and inductive measurements, except  near the front wave where the small and fast $S_H$ forerunners are not resolved by the inductive method (see inset of Fig.\ \ref{fig:formeBF}(b)).

\begin{figure}[ht!]
\epsfig{file=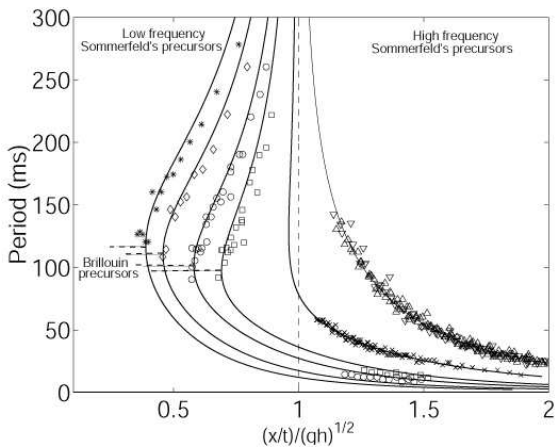, width=8.5 cm}
\caption{\label{fig:period} 
Period of Sommerfeld precursors $S_{H}$ and $S_{L}$ as a function of $(x/t)/\sqrt{gh}$ for  various heights $h=$ 2.12 for depression ($\bigtriangledown$) or elevation ($\bigtriangleup$) pulses, 3.4 ($\times$),  5.6 ($\square$), 7.2($\circ$), 10.4 ($\diamond$), 13.75 ($\ast$) mm with $0.2\le x \le 0.8$ m. For each value of $h$, the theoretical Sommerfeld ($-$) and Brillouin ($--$) precursor periods are extracted from Eq.\ (\ref{disp}) (see text for details).}
\end{figure}

Finally, the period of each forerunner is measured all along its propagation ($0.2\le x \le 0.8$ m) by recording the time between successive maxima of the amplitude. The periods are displayed in Fig.\ \ref{fig:period} as a function of $t_0/t$ ($t_0 \equiv x/\sqrt{gh}$) for various fluid depths $h$ corresponding to $0.02 \leq {\rm Bo} \leq 0.67$. For each height corresponding to ${\rm Bo} \geq1/3$, the $S_H$ precursor period increases as the waves propagate. For $0\leq {\rm Bo} < 1/3$, both $S_H$ and $S_L$ precursors are observed, and the period of $S_L$ decreases with time.  We can map Fig.\ \ref{fig:period} on Fig.\ \ref{fig:vitgroup} with a $90^{\rm o}$ rotation and $(x/t)/\sqrt{gh}$ can then roughly be viewed as the group velocity curve of the stationary mode $k_s$ as a function of $1/k_s$. Therefore, Fig.\ \ref{fig:period} shows that $S_H$ precursor velocities are supersonic ($v_g(k_{s_H})\equiv x/t>\sqrt{gh}$), whereas $S_L$ precursor is subsonic ($v_g(k_{s_L})<\sqrt{gh}$).
For each $h$, all the data recorded at different $x$, lie on a single curve predicted by Eq.\ (\ref{disp}), which is the parametric plot of $2\pi/\omega(k)$ as a function of $v_g(k)\equiv d\omega/dk$ for various values of $k$.
Note that a Brillouin forerunner is never observed in our experiments, and $S_H$ disappear for large $h$ (absence of $*$ and $\diamond$-marks on Fig.\ \ref{fig:period}). Note also the absence of $S_H$ for $(x/t)/\sqrt{gh} < 1$. Although they are predicted from Fig.\ \ref{fig:vitgroup}, they are much smaller than $S_L$, and thus cannot be observed when they travel at the same velocity.   

In conclusion, we have reported the observation of Sommerfeld forerunners in the shallow water limit of surface waves  and found that their period is in good agreement with theoretical predictions. An extension of this work, much easier to study in the context of hydrodynamics than in the one of electromagnetism, is to understand how the dynamics of precursors are changed when the main signal amplitude is increased such that nonlinear effects become important.

\begin{acknowledgments}       
We thank B. Castaing for discussions. This work has been supported by the French Ministry of Research under Grant ACI Jeunes Chercheurs 2001 and by the GDR  ``Ph\'enom\`enes hors \'equilibre''  of CNRS.
\end{acknowledgments}



\begin{thebibliography}{50}
\bibitem{Brillouin60}L. Brillouin, {\em Wave Propagation And Group Velocity}, Academic Press, Inc., Ny (1960) notably including the english translation of A. Sommerfeld, Ann. Physik (Leipzig) {\bf 44}, 177 (1914); L. Brillouin, {\em ibid}. {\bf 44}, 203 (1914). 
\bibitem{Pleshko69}P. Pleshko and I. Pal\'ocz, Phys. Rev. Lett. {\bf 22}, 1201 (1969); D.~D. Stancil, J. App. Phys {\bf 53},  2658 (1982). 
\bibitem{Aaviksoo91}J.~Aaviksoo, J.~Kuhl and K.~Ploog, Phys. Rev. A {\bf 44}, R5353 (1991)
\bibitem{Albanese89}R. Albanese, J. Penn and R. Medina, J. Opt. Soc. Am. A {\bf 6}, 1441--1446 (1989).
\bibitem{Hanyga02}A. Hanyga, Pure Appl. Geophys. {\bf 159}, 1749 (2002)
\bibitem{Mojahedi00}M. Mojahedi, E. Schamiloglu, F. Hegeler and K.~J. Malloy, Phys. Rev. E {\bf 62}, 5758 (2000); A.~P. Barbero, H.~E. Hernandez-Figueroa and E. Recami, Phys. Rev. E {\bf 62}, 8628 (2000); S. Chu and S. Wong, Phys. Rev. Lett. {\bf 48}, 738 (1982), D. Mugnai, A. Ranfagni and R. Ruggeri, Phys. Rev. Lett. {\bf 84} 4830 (2000), L.~J. Wang, A. Kuzmich and A. Dogariu, Nature {\bf 406}, 277 (2000).
\bibitem{Varoquaux86}E. Varoquaux, G.~A. Williams and O. Avenel, Phys. Rev. B {\bf 34}, 7617 (1986) and references therein.
\bibitem{Kececioglu81}I. Kececioglu, M.~E. McClurken, R.~D. Kamm and A.~H. Shapiro, J. Fluid Mech. {\bf 109}, 367 (1981); P. Flaud, D. Geiger and C. Oddou, J. Physique {\bf 47}, 773 (1986); T.~B. Moodie and J.~B. Haddow, J. Acoust. Soc. Am. {\bf 67}, 446 (1980) 
\bibitem{anormale}The dispersion is called anomalous when the group velocity exceeds the phase velocity.
\bibitem{Kranzer59}J.~E.~Prins, Trans. Am. Geophys. Union {\bf 39}, 865 (1958) for experiments; H.~C.~Kranzer and J.~B.~Keller, J. Appl. Phys. {\bf 30}, 398 (1959); H.~Lamb, {\em Hydrodynamics}, Dover, NY (1945), for Cauchy-Poisson problem of water waves generated by sudden disturbances of the free surface.
\bibitem{Smith95}M.~S.~Smith and J.~B.~Sheperd, Natural Hazards {\bf 11}, 75 (1994).
\bibitem{Handbook}Handbook of Chemistry and Physics, D~R. Lide Ed., CRC Press, USA, 80th Ed. (1999).
\bibitem{Falcon02}E.~Falcon, C.~Laroche and S.~Fauve, Phys. Rev. Lett. {\bf 89}, 204501 (2002)
\bibitem{Jeffreys56}H.~Jeffreys \& B.~Jeffreys, {\em Methods of Mathematical Physics}, Cambridge U. P., 3rd ed., Cambridge (1956), pp.498-518; T.~H.~Havelock, {\em The Propagation of Disturbances in Dispersive Media}, Cambridge U. P., Cambridge (1914); J.~J.~Stoker, {\em Water Waves}, Interscience Publishers, Inc., NY (1957); V.~I.~Karpman, {\em Nonlinear Waves in Dispersive Media}, Pergamon Press, New York (1975).
\bibitem{Whitham74}G.~B.~Whitham, {\em Linear and Nonlinear Wave}, John Whiley \& Sons,  Inc (1974), J. D. Jackson, {\em Classical Electrodynamics}, John Whiley \& Sons, 3rd ed., NY (1998).
\end{thebibliography}
\end{document}